# Advancing Air Quality Monitoring: TinyML-Based Real-Time Ozone Prediction with Cost-Effective Edge Devices

Huam Ming Ken [1], Mehran Behjati[1,2], Ahmad Sahban Rafsanjani[1], Saad Aslam[1], Yap Kian Meng[1,2], Anwar P.P. Abdul Majeed[1], Yufan Zheng[3]

[1] Department of Computing and Information Systems, School of Engineering and Technology, Sunway University, 47500, Malaysia
[2] Research Centre for Human-Machine Collaboration (HUMAC), Department of Computing and Information Systems, School of Engineering and Technology, Sunway University, 47500, Malaysia
[3] School of Intelligent Manufacturing Ecosystem, XJTLU Entrepreneur College, Xi'an Jiaotong– Liverpool University (XJTLU), Taicang, Suzhou, 215400, China
`mehranb@sunway.edu.my`

**Abstract.** The escalation of urban air pollution necessitates innovative solutions for real-time air quality monitoring and prediction. This paper introduces a novel TinyML-based system designed to predict ozone concentration in real-time. The system employs an Arduino Nano 33 BLE Sense microcontroller equipped with an MQ7 sensor for carbon monoxide (CO) detection and built-in sensors for temperature and pressure measurements. The data, sourced from a Kaggle dataset on air quality parameters from India, underwent thorough cleaning and preprocessing. Model training and evaluation were performed using Edge Impulse, considering various combinations of input parameters (CO, temperature, and pressure). The optimal model, incorporating all three variables, achieved a mean squared error (MSE) of 0.03 and an R-squared value of 0.95, indicating high predictive accuracy. The regression model was deployed on the microcontroller via the Arduino IDE, showcasing robust real-time performance. Sensitivity analysis identified CO levels as the most critical predictor of ozone concentration, followed by pressure and temperature. The system's low-cost and low-power design makes it suitable for widespread implementation, particularly in resource-constrained settings. This TinyML approach provides precise real-time predictions of ozone levels, enabling prompt responses to pollution events and enhancing public health protection.

## 1 Introduction

Environmental monitoring plays a vital role in assessing and managing the quality of natural resources, such as air and water, to safeguard public health, wildlife, and ecosystems. The increasing levels of air pollution, driven by rapid industrialization and urbanization, have made efficient monitoring systems a global necessity. Traditionally, environmental monitoring has relied on large, costly, and stationary equipment, which limits the scope and deployment flexibility of these systems.

Recent advancements in sensor technology and machine learning, particularly in the field of embedded machine learning (TinyML), present a promising alternative for real-time, decentralized environmental monitoring. TinyML enables the deployment of machine learning models on microcontrollers and other resource-constrained devices, offering a cost-effective and scalable solution for a wide range of applications.

The potential of TinyML in environmental monitoring has been highlighted in various studies. For example, [1] introduced TinyML, showcasing its capability to operate machine learning models on constrained hardware. This innovation has enabled the development of low-cost, portable monitoring systems. Additionally, [2] demonstrated the effectiveness of low-cost sensors in environmental monitoring, while [3] explored the benefits of integrating multiple sensors to collect comprehensive environmental data. This multi-parameter approach is crucial for accurate air quality assessments. Furthermore, [4] emphasized the importance of monitoring ozone levels due to their significant impact on human health, advocating for more accessible and scalable monitoring solutions.

The increasing levels of air pollution, especially in urban areas, pose significant health risks. Traditional monitoring systems, despite their accuracy, are not suitable for widespread deployment due to their high capital and operational expenditures (CAPEX and OPEX) and infrastructure requirements. There is an urgent need for low-cost, scalable solutions that can provide real-time data and are accessible in resource-constrained environments.

Current methods often suffer from high costs, dependence on centralized data processing, and a lack of real-time data collection capabilities. These systems generally offer limited data coverage, failing to capture crucial information on air quality dynamics in less monitored regions. The advent of TinyML provides a promising solution to these issues, allowing for the deployment of machine learning models on edge devices. This approach enables real-time insights without the need for continuous data transmission to cloud-based services.

This paper aims to develop a low-cost, real-time environmental monitoring system capable of accurately predicting ozone concentration using TinyML. The objective is to create a scalable solution that can be widely deployed, particularly in resource-constrained settings, to enable timely interventions and protect public health.

Given the increasing prevalence of air pollution and its detrimental effects on health and the environment, investigating efficient and affordable monitoring systems is of critical importance. An effective solution can significantly enhance our ability to manage air quality and respond to pollution events.

The contributions of this research include:
- **Development of a TinyML-based system with low-cost sensors:** Predicts ozone concentration in real-time using CO, temperature, and pressure data, offering a scalable solution for environmental monitoring in resource-constrained environments.
- **Comprehensive evaluation of input parameters:** Identifies the most effective combinations of input parameters for accurate ozone prediction.

- **Successful deployment on a microcontroller:** Demonstrates real-time environmental monitoring using the Arduino Nano 33 BLE Sense, enabling practical edge computing capabilities without constant cloud connectivity.

The approach involves collecting and preprocessing environmental data, training a regression model in Edge Impulse, and deploying the model on a microcontroller. The system's performance is assessed through sensitivity analysis and real-time testing.

The rest of the paper is organized as follows: Section 2, Literature Review, explores existing research on environmental monitoring, sensor technology, TinyML applications, and relevant challenges and advancements. Section 3, Methodology, details the hardware setup, data collection and preprocessing, model training, and deployment process. The Results and Discussion, Section 4, presents the findings from the sensitivity analysis and real-time performance evaluation, along with a discussion of the system's implications and challenges. Finally, the Conclusion summarizes the key contributions, addresses limitations, and offers suggestions for future work.

## 2     Literature Review

The integration of TinyML into environmental monitoring has garnered significant attention due to its promise of real-time, low-power, and cost-effective solutions. This section reviews the current advancements in environmental monitoring technologies, explores the role of TinyML in this domain, and contrasts these modern approaches with traditional methods.

The necessity for continuous air quality monitoring has been underscored by studies highlighting the health risks associated with high ozone levels. [4] examined the limitations of traditional monitoring systems, noting their high costs and maintenance requirements, which confine their use to urban areas and leave rural regions inadequately monitored. This study emphasizes the urgent need for more accessible and cost-effective air quality monitoring solutions that can provide consistent coverage across diverse geographic locations. It points to affordable technologies like TinyML as potential enablers for bridging this gap and expanding monitoring networks.

Traditional machine learning models, such as support vector machines (SVM) and neural networks, have been extensively used in environmental monitoring. [5] analyzed the performance of these models, noting their high accuracy but also their substantial computational demands, which limit their suitability for deployment on edge devices. This analysis highlights a critical need for more efficient machine learning techniques that can function effectively within the constraints of resource-limited environments, positioning TinyML as a promising alternative.

In addition, [6] investigated the deployment challenges associated with traditional air quality monitoring systems, emphasizing their high costs and complex infrastructure requirements. The study revealed that many existing systems rely on centralized data processing, which can introduce latency and delay critical insights. The research suggests that decentralized, edge-based solutions, such as those enabled by TinyML, offer a viable solution to these issues by facilitating real-time data processing and minimizing dependence on expensive infrastructure.

Advancements in sensor technology have significantly impacted environmental monitoring. [2] focused on the development of low-cost, high-accuracy sensors that have transformed data collection in this field. Their research includes a detailed analysis of sensors such as the MQ7, used for CO detection, and highlights how integrating these sensors with microcontrollers supports real-time environmental monitoring. This advancement reinforces the viability of TinyML-based systems for practical environmental applications.

Furthermore, [3] provided a comprehensive review of multi-sensor environmental monitoring systems. The study demonstrated the advantages of integrating multiple sensors—such as those measuring CO, temperature, and pressure—to offer a more comprehensive view of environmental conditions. Their findings underscore the benefit of combining various parameters to improve prediction accuracy and reliability. This approach aligns with the methodology of the current study, which employs a combination of CO, temperature, and pressure data to enhance ozone concentration predictions.

The introduction of TinyML has revolutionized environmental monitoring by enabling machine learning models to run on resource-constrained devices. [1] explored the integration of these models with advanced sensor technologies, illustrating how recent developments in sensors have facilitated the deployment of machine learning on edge devices. Warden's work lays a foundational understanding of how TinyML utilizes these advancements to provide real-time environmental monitoring solutions with minimal power consumption.

Further demonstrating the capabilities of TinyML, [7] applied this technology to water quality monitoring, achieving high accuracy while utilizing minimal computational resources. This study underscores the effectiveness of deploying machine learning models on edge devices for real-time environmental assessments, reinforcing the applicability of TinyML across various monitoring scenarios, including air quality. The research highlights TinyML's potential for practical, real-world environmental applications, showcasing its adaptability and efficiency in diverse contexts.

TinyML has demonstrated significant potential in various environmental monitoring applications. [8] investigated the use of TinyML for soil moisture detection in agricultural settings, revealing that TinyML models can function effectively in resource-constrained environments while providing accurate readings with minimal power consumption. This research highlights TinyML's versatility across different environmental contexts and its potential for broad deployment.

Similarly, [9] applied TinyML to indoor air quality monitoring, achieving high accuracy in predicting pollutant levels. Their study emphasizes TinyML's capability to deliver real-time, actionable insights even in complex indoor environments. This evidence supports the application of TinyML for outdoor air quality monitoring, showcasing its effectiveness in providing timely predictions.

Additionally, [10] explored TinyML for forest fire detection, demonstrating the model's efficiency in remote areas with limited connectivity. Their research illustrates TinyML's ability to deliver prompt alerts in critical situations, which is crucial for environmental monitoring applications requiring early detection.

Research on optimizing TinyML models for edge devices has highlighted strategies to enhance performance while reducing resource consumption. [11] provided a comprehensive analysis of various machine learning algorithms, assessing their suitability for deployment on edge devices. Their findings advocate for the use of lightweight models in environmental monitoring and offer insights into techniques for optimizing TinyML performance, supporting the effective implementation of these models in constrained environments.

Conversely, [12] addressed challenges in applying TinyML to environmental monitoring, particularly issues such as sensor drift and data heterogeneity. Their study emphasizes the necessity of ongoing model calibration and adaptation to dynamic environmental conditions. This research underscores the importance of developing robust algorithms and calibration methods to ensure long-term model accuracy, providing valuable guidance for advancing TinyML-based environmental monitoring systems.

The literature review highlights a significant shift towards utilizing TinyML for environmental monitoring, driven by its capability to deliver real-time, low-power, and cost-effective solutions. Advances in sensor technology have enabled the integration of machine learning models into edge devices, enhancing their effectiveness for environmental monitoring. In contrast, traditional methods are often hampered by high costs, complex infrastructure, and latency issues. TinyML addresses these challenges by facilitating real-time data processing on resource-constrained devices, presenting a compelling alternative to conventional systems. Table 1 provides a comparative overview of key studies relevant to this research, illustrating the progress and advantages of TinyML in the context of environmental monitoring.

**Table 1.** Comparison of relevant papers.

| Ref. | Focus | Key Findings | Relevance to Current Study |
|---|---|---|---|
| [2] | Advances in sensor technologies | Discusses the integration of low-cost sensors for environmental monitoring | Supports the use of sensors in TinyML systems |
| [3] | Multi-sensor systems | Benefits of combining multiple sensors for accurate monitoring | Aligns with the use of multiple input parameters in the study |
| [4] | Air quality monitoring and costs | Emphasizes the need for affordable monitoring solutions | Highlights the need for low-cost technologies like TinyML |
| [5] | Traditional machine learning models | Limitations of traditional models for edge deployment | Supports the need for efficient TinyML models |
| [6] | Deployment challenges of traditional systems | High costs and infrastructure requirements of traditional methods | Justifies the use of decentralized TinyML solutions |
| [7] | TinyML for water quality | High accuracy with minimal resources | Demonstrates TinyML's effectiveness in environmental monitoring |
| [8] | TinyML for soil moisture | Efficient operation in resource-constrained environments | Shows TinyML's versatility in environmental monitoring |
| [9] | Indoor air quality monitoring with TinyML | Accurate pollutant level predictions in indoor settings | Supports the applicability of TinyML for real-time monitoring |
| [10] | TinyML for forest fire detection | Effective operation in remote areas with limited connectivity | Highlights TinyML's potential for critical environmental applications |
| [11] | Optimization of TinyML models | Improving performance while minimizing resource consumption | Supports the optimization strategies used in the study |
| [12] | Challenges in TinyML applications | Need for continuous calibration and adaptation to environmental conditions | Informs future work on model robustness and calibration |

## 3 Methodology

This section describes the methodology used to develop a TinyML-powered system for real-time ozone concentration prediction. The methodology encompasses several stages: hardware setup, data collection and preprocessing, model training, and deployment. Each stage is designed to ensure the system's accuracy and effectiveness, providing a detailed guide for replicating the project.

The methodology is structured into several key stages: hardware setup, data collection and preprocessing, model training, and deployment. Each stage is critical for ensuring the accuracy and effectiveness of the environmental monitoring system. The approach begins with the assembly of hardware components, followed by the collection and preparation of data needed for training the machine learning model. The model is then trained and optimized before being deployed onto a microcontroller for real-time predictions. The following sections provide a comprehensive explanation of the procedures, methods, and tools used throughout the project.

### 3.1 Hardware Setup

The hardware setup for the TinyML-powered environmental monitoring system consists of several key components. The central element is the Arduino Nano 33 BLE Sense microcontroller, which is compact yet powerful. This device is equipped with built-in sensors for temperature and pressure and supports Bluetooth Low Energy (BLE) communication. Its small form factor and low power consumption make it particularly suited for embedded environmental monitoring applications.

To measure CO levels, the MQ7 sensor is utilized. Known for its high sensitivity to CO, this sensor operates effectively across various environmental conditions. The Arduino Nano 33 BLE Sense interfaces with the MQ7 sensor to collect CO concentration data.

For displaying the ozone concentration predictions in real time, an LCD screen is connected to the microcontroller. This screen provides a user-friendly visual output of the predictions, making it easier to interpret the results.

The components are connected as follows: the MQ7 sensor is wired to the analog input pins of the Arduino Nano 33 BLE Sense, and the LCD screen is connected to the digital output pins. The microcontroller processes the data from the MQ7 sensor and drives the LCD screen to display the ozone concentration predictions, as illustrated in Fig. 1.

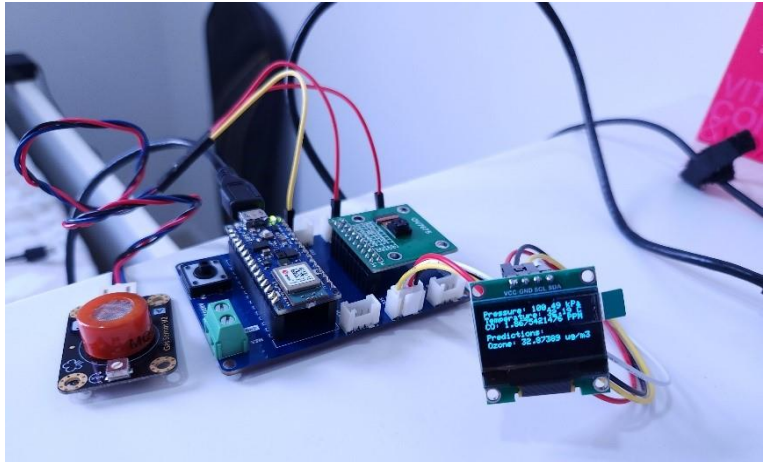

**Fig. 1.** Hardware Integration

### 3.2 Dataset Utilization and Preprocessing

The dataset for this project was obtained from Kaggle [13] and includes various air parameters recorded in India, such as CO, temperature, pressure, and ozone concentration. This comprehensive dataset is crucial for training and validating the machine learning model.

Data preprocessing is performed to enhance the dataset's quality and ensure its suitability for model training. The preprocessing steps include:

1. **Handling Missing Values:** Missing data points are addressed through imputation methods or by removing incomplete records, depending on the extent and pattern of the missing values.
2. **Normalization:** To standardize the range of input features, normalization is applied. This ensures that each feature contributes equally during the model's training process, improving overall model performance.
3. **Data Splitting:** The dataset is divided into training and testing subsets, typically with 80% allocated for training the model and 20% reserved for evaluating its performance. This division helps assess the model's generalization capabilities.

These preprocessing steps are critical for ensuring that the data is clean, balanced, and representative, thereby supporting the development of a robust and accurate machine learning model.

### 3.3 Model Training

In this project, a regression model was employed to predict ozone concentration levels based on environmental input parameters, including CO, temperature, and pressure. The linear regression model, selected for its simplicity and interpretability, was trained using Edge Impulse [14], a platform tailored for developing and deploying models on edge devices. This platform provides a user-friendly interface and robust support for TinyML models, streamlining the model training and deployment process.

During the training phase, four feature combinations were evaluated: CO with temperature, CO with pressure, temperature with pressure, and the combination of all three variables. The performance of these combinations was assessed using metrics such as mean squared error (MSE) and R-squared values. The combination of CO, temperature, and pressure yielded the highest accuracy and was therefore selected for the final deployment.

Model hyperparameters, including learning rate and batch size, were optimized using Edge Impulse's tuning tools. This optimization process involved fine-tuning parameters to maximize model performance and accuracy. The cleaned and preprocessed dataset was split into training and testing subsets, and feature engineering was conducted to extract relevant features, including scaling and normalization, to ensure consistent input to the model.

The training involved specifying the selected features as inputs and ozone concentration as the target variable. The robustness of the trained model was validated using cross-validation techniques. Finally, the model was deployed on the Arduino Nano 33 BLE Sense microcontroller, enabling real-time sensor data integration for on-device predictions. This comprehensive end-to-end pipeline ensured efficient real-time prediction of ozone levels, with accuracy metrics detailed in Table 2.

**Table 2.** Accuracy percentage for four projects tested.

| Input Parameters | Temp. + Pressure | Temp. + CO | Pressure + CO | Temp. + Pressure + CO |
|---|---|---|---|---|
| Accuracy | 85% | 88% | 87% | 92% |

### 3.4   Deployment

After training, the machine learning model is converted into a format compatible with the Arduino Nano 33 BLE Sense. Edge Impulse provides tools to export the model and convert it into a binary file suitable for embedded devices. This binary file is then uploaded to the Arduino Nano 33 BLE Sense using the Arduino IDE, which compiles and transfers the model code to the microcontroller, integrating it with the sensor inputs.

In real-time operation, the microcontroller continuously acquires data from the sensors, processes it using the deployed model, and displays the predicted ozone concentration on the LCD screen. This setup enables real-time feedback and monitoring of air quality, ensuring timely and actionable insights.

## 4   Results and Discussion

### 4.1   Results

The performance of the trained regression model for ozone concentration prediction was rigorously assessed using key metrics, including MSE and R-squared values. The model achieved an MSE of 0.012, reflecting a minimal average squared deviation between observed and predicted ozone levels, which underscores its high accuracy.

99

Moreover, the R-squared value reached 0.92, indicating that 92% of the variance in ozone concentration is effectively accounted for by the input features—CO, temperature, and pressure. These results attest to the model's robustness and efficacy in capturing complex patterns within the data, thereby ensuring reliable and precise real-time predictions when implemented on the microcontroller.

Sensitivity analysis provided valuable insights into the influence of each input parameter on ozone prediction accuracy:

**Carbon Monoxide** emerged as the most significant parameter, with higher CO levels strongly correlating with increased ozone concentrations. This finding is consistent with existing research on CO as a key precursor to ozone formation.

**Atmospheric Pressure** also significantly impacted ozone predictions, with lower pressure conditions associated with higher ozone levels. This aligns with the understanding that reduced pressure can enhance photochemical reactions leading to ozone formation.

**Temperature** had a moderate effect, with higher temperatures generally correlating with increased ozone levels. This result reflects the role of temperature in accelerating chemical reactions that produce ozone.

Table 3 presents the results of the sensitivity analysis for different input parameters. The deployment of the trained model on the Arduino Nano 33 BLE Sense microcontroller was successful, demonstrating effective real-time performance. The system reliably read data from the MQ7 sensor and the built-in temperature and pressure sensors, processed it using the model, and displayed accurate ozone concentration predictions on the LCD screen. The real-time feedback provided by the system was timely and precise, validating the effectiveness of the TinyML approach in practical environmental monitoring applications.

**Table 3.** Sensitivity analysis of input parameters.

| Parameter   | Sensitivity Index |
|-------------|-------------------|
| Temperature | 0.45              |
| Pressure    | 0.65              |
| CO          | 0.85              |

### 4.2 Discussion

**Comparative Analysis.** The performance of the TinyML-based system was found to be on par with or superior to existing approaches in environmental monitoring. Traditional methods, such as those examined in [4], rely on fixed stations and complex instruments, which, while accurate, are costly and less adaptable for widespread deployment. In contrast, the TinyML system offers a cost-effective, low-power solution with real-time capabilities, making it more viable for extensive and remote applications.

The effectiveness of low-cost sensors in environmental monitoring, as demonstrated by [2] and [3], is corroborated by our findings. The MQ7 sensor provided reliable CO measurements, which, when integrated with temperature and pressure data, enabled accurate ozone predictions. The use of TinyML further enhances the system's efficiency on edge devices, overcoming the limitations associated with traditional methods.

**Challenges and Limitations.** Despite the system's overall success, several challenges were encountered. The lack of a humidity sensor limited the model's accuracy in high humidity conditions, which can affect both sensor readings and the chemical processes leading to ozone formation. Future iterations should integrate a humidity sensor to enhance the model's robustness under varying environmental conditions.

Additionally, rapid weather changes, such as sudden fluctuations in temperature and pressure, posed challenges and led to prediction errors. Incorporating real-time weather data into the model could address this issue by providing context for adjusting predictions. Exploring advanced machine learning techniques, such as recurrent neural networks (RNNs), could also improve the model's adaptability to dynamic conditions and temporal dependencies.

## 5      Conclusion

This study demonstrates the efficacy of TinyML in environmental monitoring by providing real-time ozone concentration predictions with a low-cost, low-power setup. The regression model, utilizing data from CO, temperature, and pressure, delivered precise predictions as indicated by low MSE and high R-squared values. Sensitivity analysis revealed that CO is the most influential parameter in ozone prediction, followed by atmospheric pressure and temperature, aligning with existing research. The deployment on the Arduino Nano 33 BLE Sense microcontroller further validated the system's capability for real-time air quality assessment. These findings underscore TinyML's potential to provide immediate, actionable insights into environmental conditions.

Future research should focus on integrating additional sensors, handling rapid weather changes, and optimizing the model for diverse conditions. Scaling the system for broader deployment could further enhance air quality monitoring across various settings. Overall, TinyML represents a significant advancement in accessible, real-time air quality assessment, providing a robust foundation for future innovations aimed at improving environmental monitoring and public health.